\documentclass[12pt,preprint]{aastex}

\begin{document}



\title{Probing the RGB-phase transition: Near--IR photometry\\
of six intermediate age LMC clusters}

\author{Francesco R. Ferraro}
\affil{Dipartimento di Astronomia, Universit\`a 
degli Studi di Bologna, Via Ranzani, 1 - 40127
Bologna, ITALY}
\email{ferraro@bo.astro.it}
\author{ Livia Origlia}
\affil{INAF - Osservatorio Astronomico di Bologna, Via Ranzani, 1 - 40127
Bologna, ITALY}
\email{origlia@bo.astro.it}
\author{Vincenzo Testa}
\affil{INAF - Osservatorio Astronomico di Roma, Via Frascati, 33 - 00040
Monteporzio Catone, ITALY}
\email{testa@mporzio.astro.it}
\and
\author{Claudia Maraston }
\affil{Max-Planck-Institut f\"ur Extraterrestrische Physik, 
Giessenbachstraße, 85748 Garching bei M\"unchen, GERMANY}
\email{maraston@mpe.mpg.de}

\altaffiltext{1}{Based on observations collected at the European Southern
Observatory, La Silla, Chile, using SOFI at the 3.5m NTT, within the observing
programs 64.N-0038 and 68.D-0287.}

\begin{abstract}
This is the first of a series of papers devoted to a global
study  of the photometric properties of the red stellar
sequences  in a complete sample of the Large Magellanic
Cloud  clusters, by means of near infrared array
photometry. Deep J,H,Ks photometry and accurate Color
Magnitude Diagrams  down to $\rm K\approx 18.5$, i.e.
$\approx 1.5$~mag  below the red He-clump,  for six
intermediate age clusters (namely NGC~1987, NGC~2108, 
NGC~2190, NGC~2209, NGC~2231, NGC~2249) are presented.  A
quantitative estimate of the population ratios (by number
and luminosity) between Red Giant Branch and He-clump stars
for each target cluster is provided and discussed in the
framework  of probing the so-called Red Giant Branch phase
transition (RGB {\it Ph-T}). 
By using the Elson \& Fall {\it s-parameter} as an age indicator, 
the observed RGB population shows a sharp enhancement
(both in number and luminosity) at $s=36$. Obviously,
the corresponding absolute age 
strictly depends on the details of theoretical models adopted  to 
calibrate  the {\it s}-parameter.  
Curiously, the currently
available calibrations of the {\it s}-parameter in term of age
based on {\it canonical} (by Elson \& Fall 1988) and
{\it overshooting} (Girardi et al. 1995) models provide ages
that well agree within $10\%$, suggesting that the full
development of the Red Giant Branch occurs at  
$t\approx 700$~Myr and be a  relatively fast event ($\delta
t\approx 300$~Myr).\\ 
However, the RGB {\it Ph-T} epoch derived from the 
{\it overshooting} calibration of the {\it s}-parameter 
turns out to be significantly earlier than the epoch  
provided by the recent evolutionary tracks by Girardi et al. (2000).
A new calibration
of the {\it s}-parameter based on high quality Color Magnitude
Diagrams and updated models is urged to address the origin of 
this discrepancy and finally establish the epoch of the RGB {\it Ph-T}.
\end{abstract}


\section{Introduction}
 
Stellar evolution theory predicts that evolved red giant stars
dominate the bolometric luminosity of a Simple Stellar Population (SSP) 
after a few  hundred million years.
Theoretical models \citep{rb86, bruzual93, maraston98}         
suggest that two special events (the so--called
{\it Phase Transitions, Ph--T}) significantly mark the spectral
evolution of a SSP during its lifetime.
The first, after $\approx 10^8$ yrs, due to the
sudden appearance of red and bright Asymptotic Giant Branch (AGB) stars 
(AGB {\it Ph--T}), and the
second, after $\approx 6\times 10^8$ yrs, due to the development of the full
Red Giant Branch (RGB {\it Ph--T}).

From an observational point of view, the globular cluster system
of the Magellanic Clouds (MC) provides an unique opportunity
to study the AGB and RGB properties (morphology, luminosity function etc.) 
with varying the age and chemical composition of the stellar population.
It has been known for a long time that MC clusters are different in many
respect from those in the Milky Way.
They represent the {\it ideal laboratory}  to investigate
the spectral behavior of a SSP since they cover a wide range in age
\citep[see e.g.][]{swb80,ef85,girardi95}, 
and metallicity \citep[see e.g.][]{sagar89}.
They also have a wide spread in integrated colors:
\citet{perss83}
showed that very young blue clusters (SWB--type I--III) do
not have extended AGB or RGB sequences, while red clusters
having SWB--type V
or later display a well populated AGB and RGB.            
Hence, SWB--type IV clusters
represent the transition class which starts to populate the
giant branches of the Color Magnitude Diagram (CMD) and whose integrated colors
become red \citep[see also Fig.1 in][ hereafter F95]{ferraro95}.
 
In the last decade high quality optical CMDs of MC cluster 
from both ground-based 
\citep[see e.g.][]{bro96,testa99,dirsch00} and 
HST \citep[see e.g.][]{ols98,bro01,rich01,mat02} 
observations down to the Turn-off have been obtained.   

However, such an optical photometric database need to be
complemented with  high quality and homogeneous IR
photometry, to  properly study the red stellar sequences. 
In this respect, we have undergoing a long-term project
devoted to study the photometric  properties of the AGB and
RGB with varying the stellar age and metallicity  in the
stellar clusters of our Galaxy and in the MC,  by coupling
the information from optical and near  IR CMDs and
Luminosity Functions (LFs) \citep{fer99,fer00}. 

Near IR observations are crucial in such a study
since they provide the highest sensitivity to the physical parameters
of cool stars and 
the contrast between red giants and
the unresolved background is greater
than in any other spectral range.
This drastically reduces the crowding effects in the innermost regions 
of the clusters and it also represents a major, conceptual simplification 
in the interpretation of the integrated spectrum.   
 
A preliminary IR survey of 12 intermediate age LMC clusters 
has been performed at the CTIO
with the first generation IRIM ($58\times64$ pixel) imager
(see F95).
Although the large pixel size and the moderate performance
of the detector strongly limited the quality of the photometry
in the crowded central region,
we sampled a good number of giants along the bright
portion (i.e. $\rm 14.3<K<12.3$)
of the RGB sequence (see Figures~5a,b in F95).
This dataset was successfully used to derive
the first direct observational evidence of the RGB {\it Ph--T} (see 
F95 for a full discussion).
However, the limited extension in magnitude of the sample
prevented us to quantify the global contribution of the RGB
to the integrated cluster bolometric light and to perform a
detailed comparison with the theoretical expectations
\citep[see Fig.~23 in][]{maraston98}.       

More recently, taking advantage of the superior
performances of the new  generation of $1024\times1024$ IR
array detectors available at  the ESO telescopes, we
performed   a photometric survey of a complete sample of
LMC clusters  of different ages. Homogeneous J,H,Ks
photometry  1.5 magnitude below the He-clump have been
obtained for 20 clusters (spanning an age range between
10$^8$ and a few 10$^9$ yrs)  during two successful
observational runs.  

This is the first of a series of papers devoted to study 
the photometric properties of the observed clusters.
In this paper, we present the results from the first
run of observations in which near IR photometry of 
6 LMC clusters with intermediate-age has been secured.
This sample represents $\approx 30\%$
of the global dataset secured within this project.

\section{Observations and photometric analysis}

J,H,Ks images of six intermediate age globular clusters in
the LMC  (cf. Table \ref{tab1}) have been obtained at ESO,
La Silla, on January 12-14, 2000, by using the  near IR
imager\-/\-spec\-trometer SOFI \citep{mcl98}  mounted at
the ESO 3.5m NTT. SOFI is an imager/spectrometer equipped
with a  $1024\times1024$ Rockwell IR-array detector; in the
imaging mode the camera can be used with two different
pixel scales: $0.292''/pixel$ and $0.145''/pixel$. All the 
observations presented here, have been performed with a
scale of $0.292''/pixel$, providing a $\approx 5'\times 5'$
field of view  each frame. Total integration times of 2 min
in J,H and 8 min in Ks  split into sets of shorter
exposures  have been achieved, allowing photometry down to 
J$\approx$19 and H,Ks$\approx$18.5  with a S/N$\ge$30.

A control field a few arcmin away from each cluster center 
has been also observed using the same instrumental
configuration,  in order to construct median-averaged sky
frames.  A large sample of high S/N flatfields in each
filter have been acquired  by using an halogen lamp,
alternatively switch on and off.  The final cluster and
control field frames are sky--subtracted and  flatfield
corrected. For each cluster, the dataset includes three
J,H,Ks images,  centered on the cluster and three
corresponding frames of an external field. 

The photometric analysis was done by using {\it daophot-II}
under IRAF.   For each observed field all the images in the
J,H,Ks filters  were carefully aligned and trimmed in order
to have three output images,  one per filter, slightly
smaller than the original ones but perfectly registered. 
Then, a first {\it daophot-II} Point Spread function (PSF) 
fitting run was applied to the J image. The output catalog
with the instrumental magnitudes has been checked for any
spurious detection or missing object  (typically 3--4 stars
at most) which have been included in the catalog  by hand.
The J catalog was then used as input {\it master}  list to 
perform PSF fitting on the H and Ks images. 

\begin{figure}
\plotone{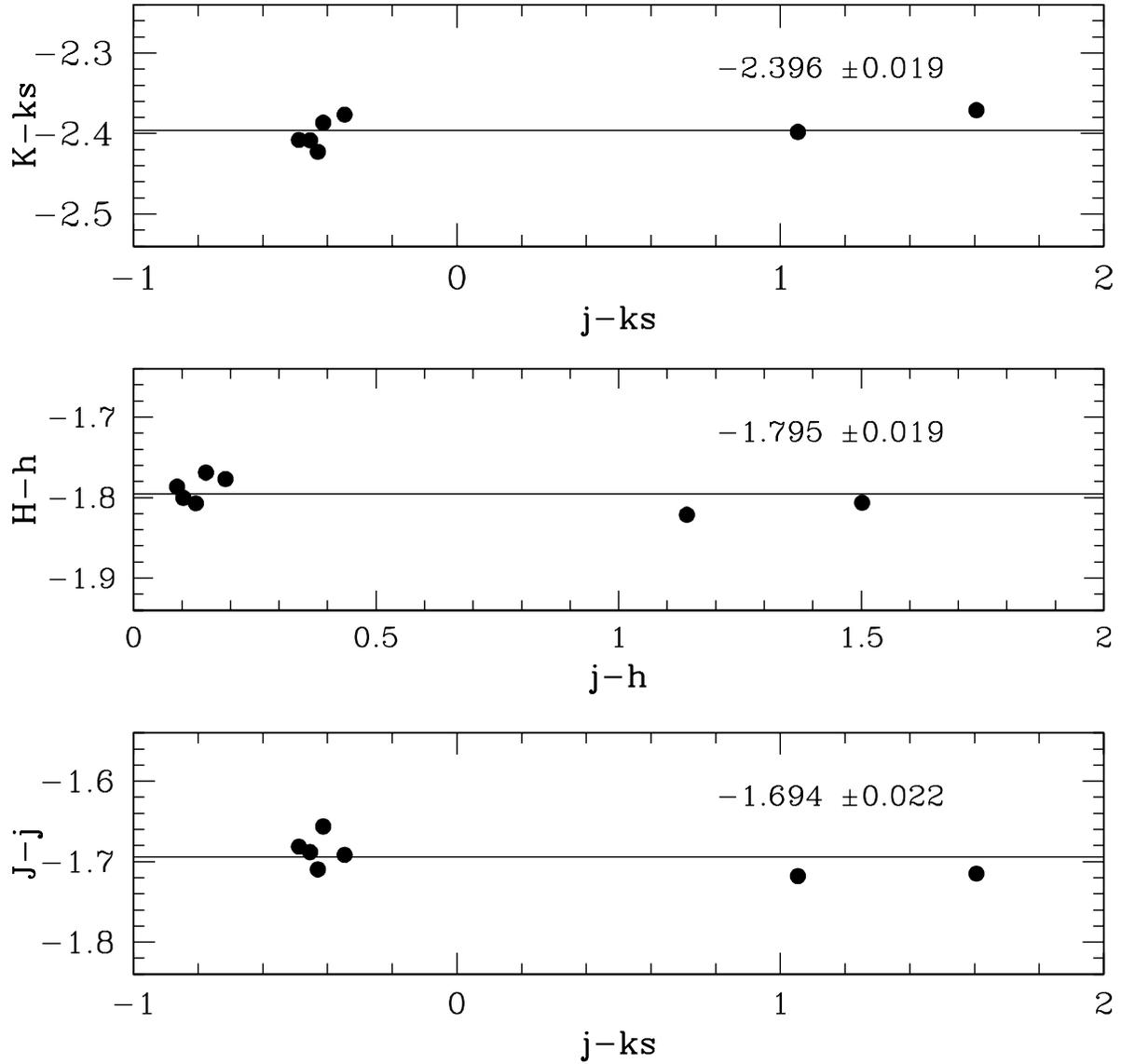}
\caption{Photometric calibration. 
$j,h,k_s$ are the instrumental aperture magnitudes 
normalized at $t_{exp}=1 sec$ and J,H,K are the corresponding 
calibrated magnitudes 
from the list of \citet{perss98}. 
\label{figure1}}
\end{figure}

A set of seven photometric standard stars spanning a wide range of 
$\rm (J-K)$ colors (about two magnitudes) 
from the list of \citet{perss98} have been 
observed during each night. 
Each standard star was observed in a sequence of five measurements  
per filter in different positions on the array.
As usual, aperture photometry has been applied to
each single frame and multiple measurements in each filter
for each standard star have been averaged together.
The final calibration of the run linking the instrumental
aperture photometry ({\it daophot-II} output) with the photometric 
standard system are as follows (see also Figure \ref{figure1}):
$$ \rm K-ks = -2.396\pm 0.019$$
$$ \rm H-h = -1.795\pm 0.019$$
$$ \rm J-j = -1.694\pm 0.022$$
where $j,h,ks$ are instrumental aperture magnitude, normalized at 
$t_{exp}=1 sec$ and J,H,K are the corresponding magnitudes 
from the list of \citet{perss98}.
Observations have been performed in the Ks filter but the calibration 
has been reported to the standard K filter. 
From the inspection of 
Figure \ref{figure1} it is clear that the color term
(if any) is very small and it can be neglected.

As usual, the most isolated brightest stars in each science
field have been used to link the aperture magnitudes to the
instrumental ones as obtained from the PSF-fitting
procedure.  The calibrated photometric catalogs in each
filter  were finally matched and merged together  in a
global catalog.  The final database consists of a table per
cluster,  listing positions and magnitudes for all the
detected objects.  Table \ref{tab2} shows an extract of the
whole database, which is entirely available in  electronic
form.

\section{Color--Magnitude and Color--Color Diagrams}
\label{cmdsec}
 
\begin{figure}
\figurenum{2}
\plotone{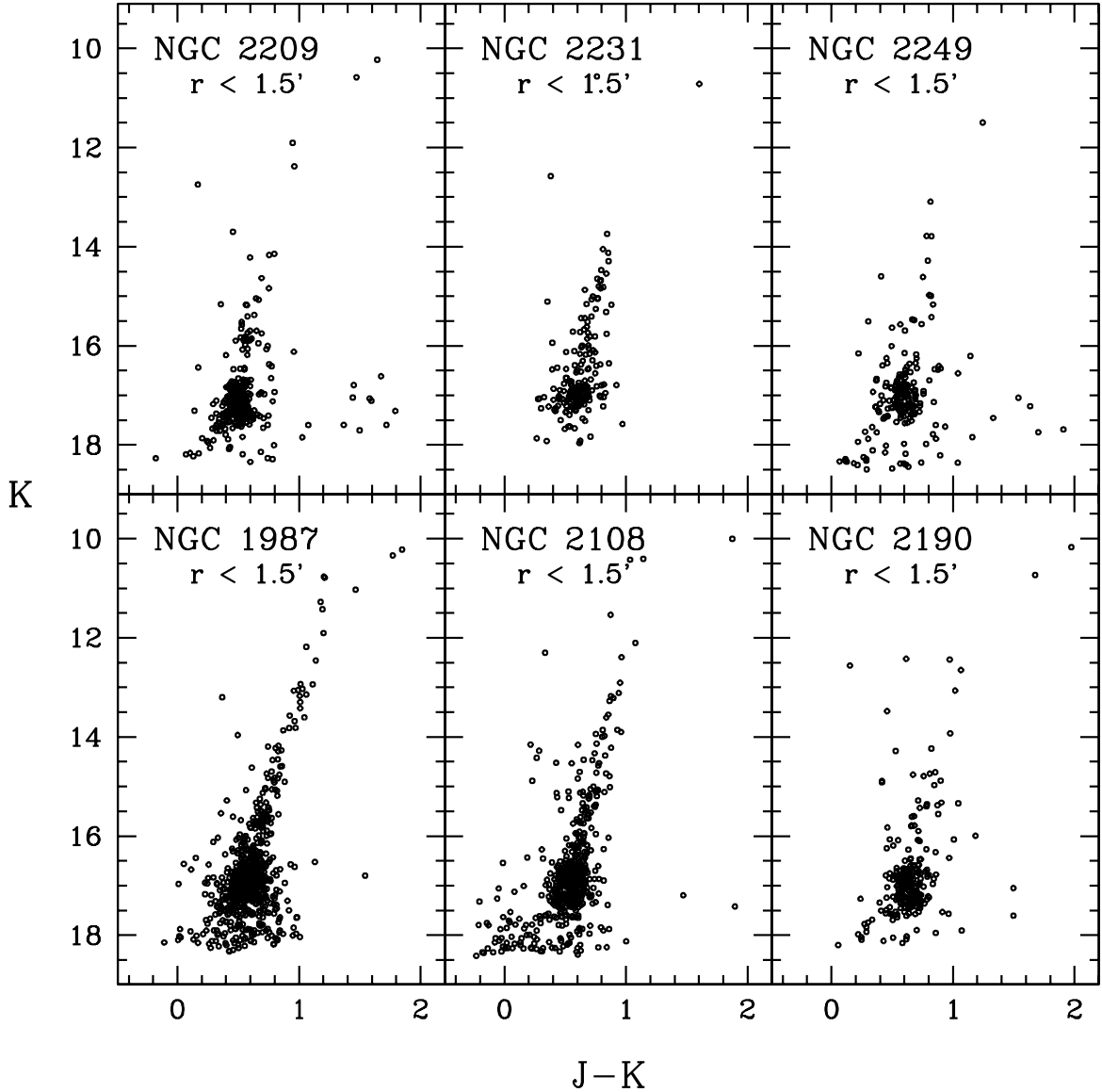}
\caption{Observed $\rm (K,J-K)$ CMDs of 
the six observed LMC clusters. Only stars at  
$r<1.5'$ from the cluster center  are plotted, 
sampling a total area of $\rm \approx~7~arcmin^2$.
 \label{figure2}}
\end{figure}

\begin{figure}
\figurenum{3}
\plotone{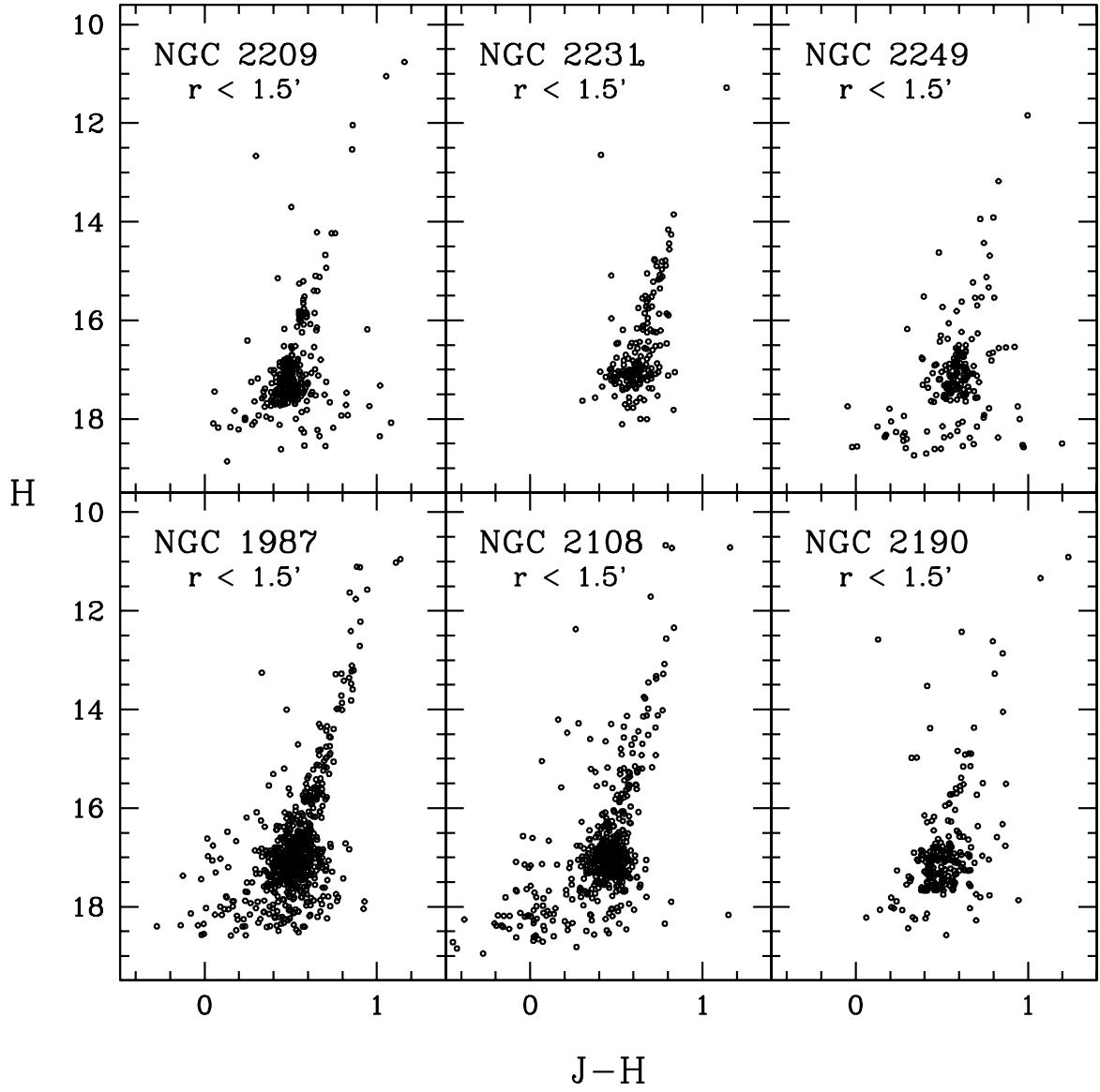}
\caption{As in figure \ref{figure2} but in the $\rm (H,J-H)$ plane.
\label{figure3}}
\end{figure}

Figures \ref{figure2} and \ref{figure3} show the CMDs for the 
observed clusters in the $\rm (K,J-K)$
and $\rm (H,J-H)$ planes, respectively.
All stars with $r <1.5'$  
from the cluster center (evaluated by eye directly on the image) 
have been plotted. 
This selection has been
done in order to minimize the field contamination and
to allow  a fist-order inspection of the cluster population.

\begin{figure}
\figurenum{4}
\plotone{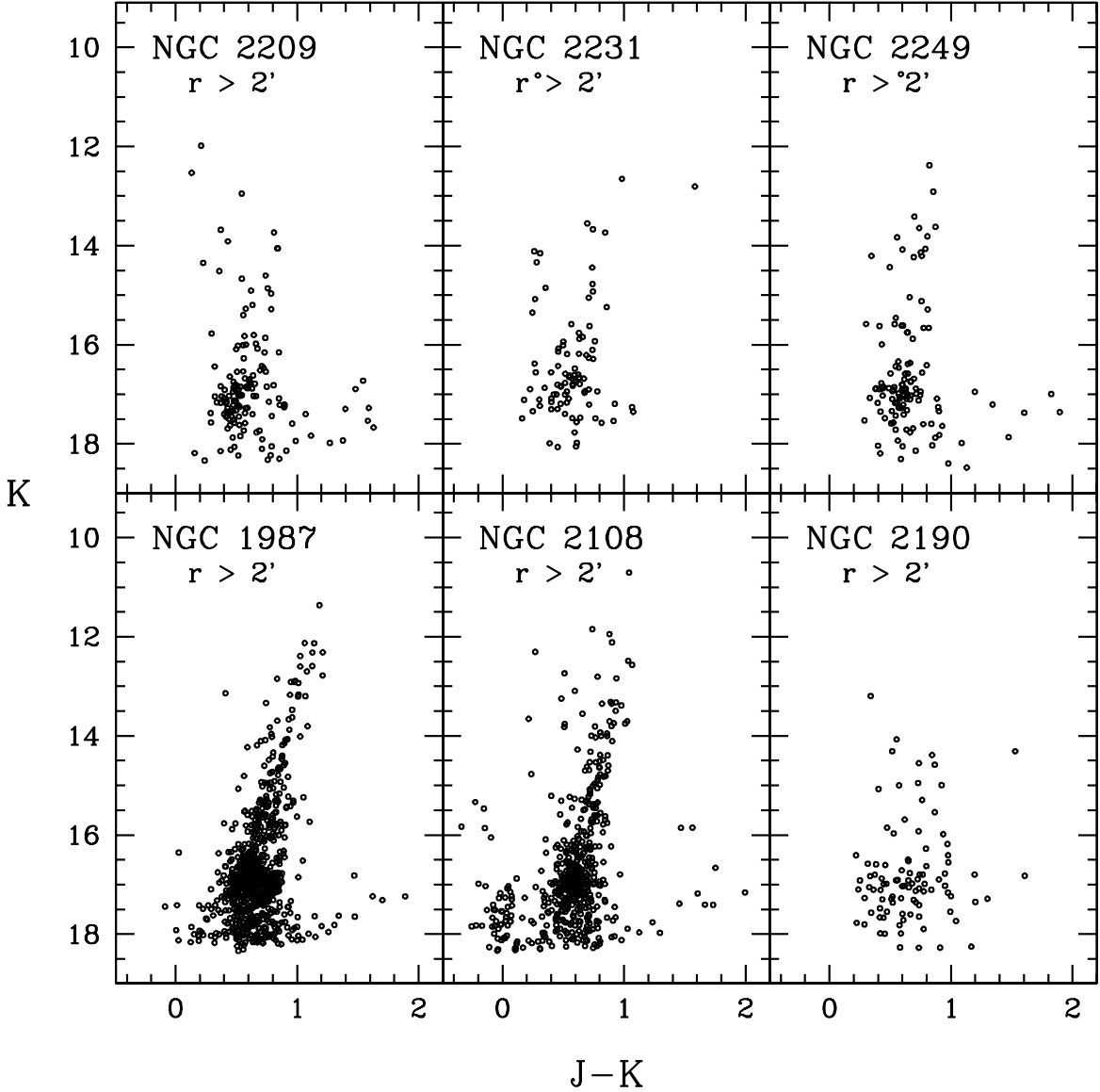}
\caption{Observed $\rm (K,J-K)$ CMDs of the outermost ($r>$2') regions
for the six observed LMC clusters. 
These CMDs are representative of the field population 
and sample an area of $\rm \approx~13~arcmin^2$.
\label{figure4}}
\end{figure}

Figure \ref{figure4} shows the $\rm (K,J-K)$ CMDs for the
external regions (at $r>2'$ from the cluster center) of the
six clusters.  These diagrams can be considered
representative of the field population. From the comparison
of Figures \ref{figure2} and \ref{figure4},  taking into
account that the field refer to an area which is double in
size  with respect to the one sampled by the {\it
contaminated} cluster population,   one can see that only
two clusters in our sample, namely  NGC~1987 and NGC~2108,
show significant field contamination. 

The main characteristics of the CMDs can be schematically summarized as
follows:

{\it (i)~}Magnitude limits down to H,K$\approx18.5$,
i.e. about 1.5~mag below the He-clump, which is
clearly visible as a concentration of stars at $\rm K\approx 17$.

{\it (ii)~}The brightest objects at $\rm K<12$ are more likely AGB stars.

{\it (iii)~}The RGB appears as a well-populated 
sequence at $\rm K>14$ in three clusters, namely NGC~1987, NGC~2108 and NGC~2231, 
although a significant fraction of stars in NGC~1987 and NGC~2108 
can belong to the field population. 
At $\rm K\lesssim 14$ very few giants are 
observed.

{\it (iv)~} In the other three clusters, namely NGC~2190, NGC~2209 and NGC~2249 only the 
red He-clump region at $\rm K\approx 17$ is well-populated, while the upper RGB 
is barely defined.

According to the historical \citet{swb80} classification
based on the {\it s}-parameter and subsequent calibrations
of such a parameter with age  \citep[see
e.g.][]{ef85,girardi95},  the LMC clusters in our sample
should have ages  in the 500-900~Myr range  (see Table
\ref{tab1} and Section \ref{compar})  and are expected to
be on the verge of the RGB {\it Ph--T} (see also F95).

In order to perform a quantitative analysis of the
population distribution  along the evolved sequences and to
compare the results  with theoretical predictions,  we
construct de-reddened CMDs and select suitable regions to 
define the AGB and RGB populations.

Correction for extinction is computed according to the $\rm
E(B-V)$  values reported in Table \ref{tab1} and the
\citet{rl85} interstellar extinction law.  The infrared
dust maps by \citet{sfd98} in the direction of  the
observed clusters provide very similar  (on average within
$\pm$0.03~dex) $\rm E(B-V)$ corrections,  with the
exception of NGC~2108,  for which the discrepancy is about
0.1~dex. However, the overall impact of such a discrepancy
on the infrared magnitudes  is always small (well within
0.1~dex), hence reddening correction is not  a critical
issue in this context.

\begin{figure}
\figurenum{5}
\plotone{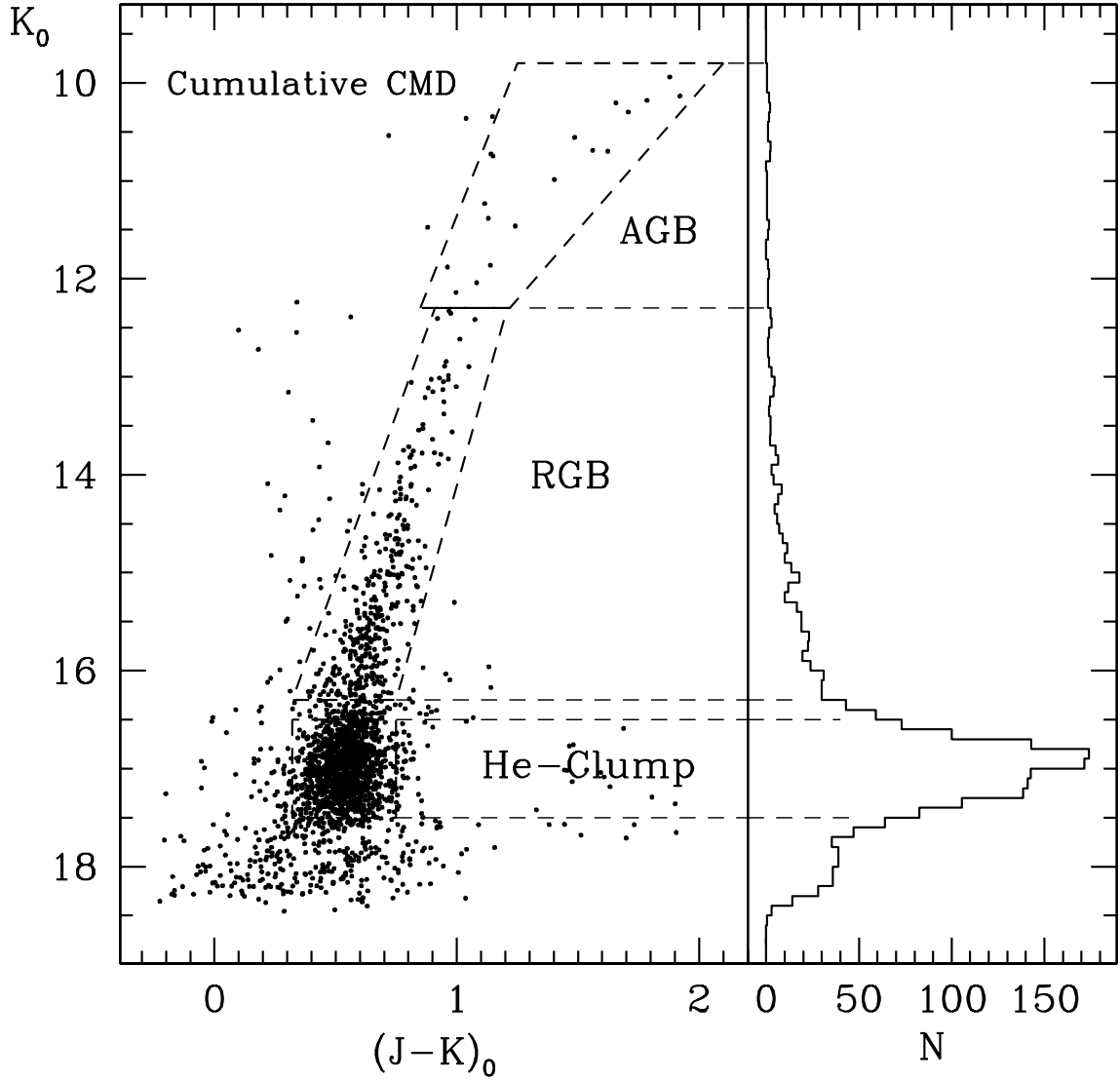}
\caption{Cumulative $\rm K_0,(J-K)_0$ de-reddened 
CMD (left panel) for the six observed LMC clusters.
The sketched regions show schematically 
the AGB, the upper RGB and red He-clump loci.
The corresponding differential luminosity function 
is also shown for clarity (right panel). 
\label{figure5}}
\end{figure}

Figure \ref{figure5} shows the cumulative $\rm
K_0,(J-K)_0$ CMD for the six observed clusters.
On this CMD we identified three {\it selection boxes}
 sampling  the bulk of the AGB, RGB and He-clump 
populations, respectively (see Figure \ref{figure5}). 
According to the observed near IR CMDs of the 10 LMC 
clusters by F95, the RGB Tip is 
expected to be at $\rm K_0\approx 12.3$.  
Recently, this value has been confirmed by computing 
the RGB LF of the LMC field 
population  from the 2MASS survey \citep{nw00}. 
From the DENIS survey \citet{cioni00} 
defined a slightly brighter (by $\approx$0.2 mag) RGB Tip 
than F95 and \citet{nw00}.  
However, since small differences in the location of the 
RGB Tip have negligible effects 
on our analysis we do not consider this issue crucial and we adopted the  
F95 and \citet{nw00} value.  
Stars brighter than the RGB Tip at $\rm K_0\approx 12.3$ and with 
$\rm (J-K)_0$ colors between 0.85 and 2.1 are thus classified as AGB.
The bulk of the bright RGB population includes all 
the stars $\approx$4 mag from the RGB Tip 
down to $\rm K_0\approx 16.3$ with $\rm (J-K)_0$ colors between 0.32  
and 1.21 (see Figure \ref{figure5}). 
The faint end of the RGB {\it selection box} has been
conservatively assumed at $\rm K_0=16.3$ (i.e. $\approx 0.7$
mag brighter than the He-clump peak at $\rm K_0=17.0$), in order to avoid a 
possible contamination  
by the He-clump distribution wings
(which are extending up to 
$\pm$0.5 mag from the peak, see right panel of Figure \ref{figure5}). 
 
\begin{figure}
\figurenum{6}
\plotone{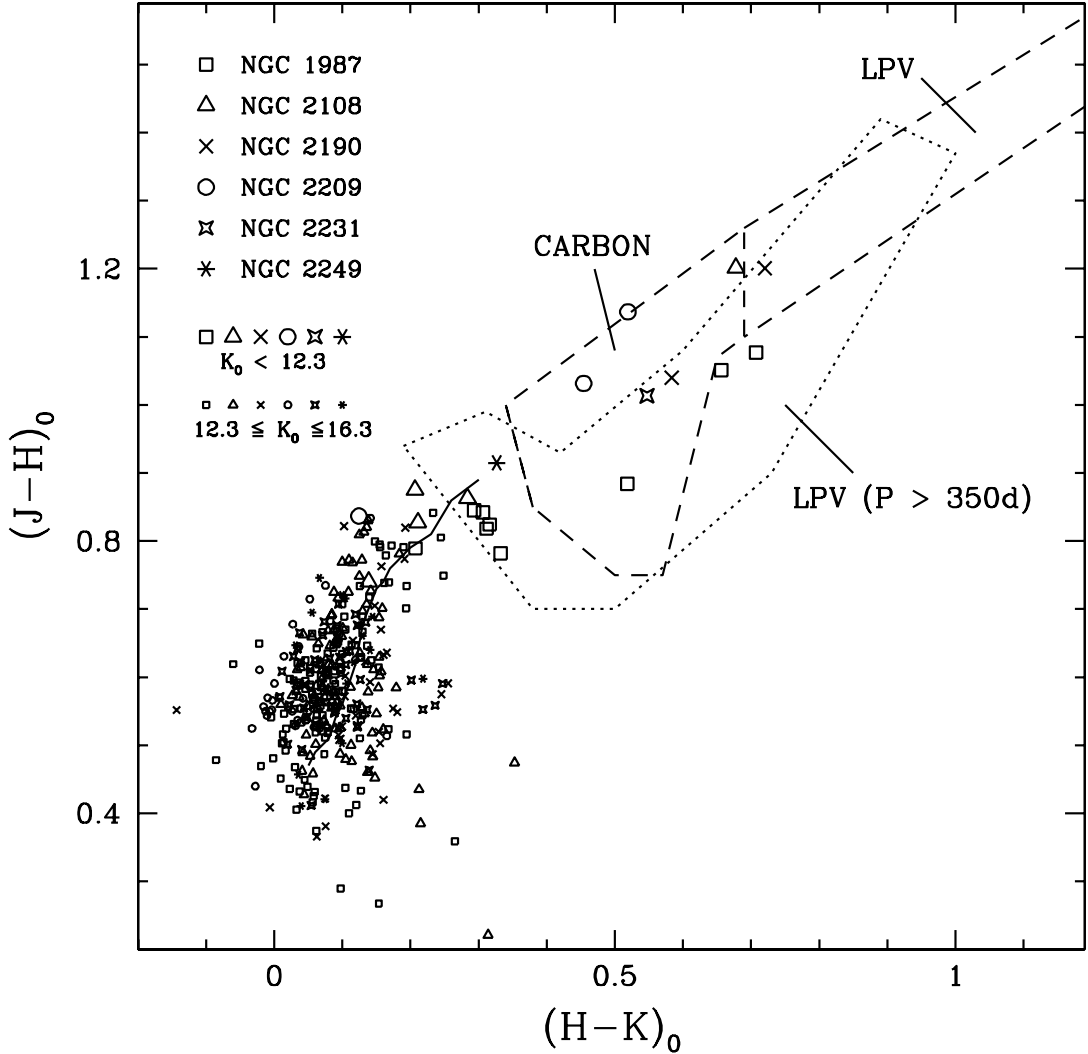}
\caption{Cumulative $\rm (J-H)_0,(H-K)_0$ de-reddened 
color-color diagram for the six observed LMC clusters.
Only the AGB and RGB stars, as classified
accordingly to the {\it selection boxes}
defined in Section \ref{cmdsec} 
and sketched in Figure \ref{figure5}, are considered.
Small symbols are stars with $\rm 16.3\le K_0 \le 12.3$ 
while big symbols represents  stars brighter than
the RGB Tip (at $\rm K_0<12.3$), and they are, most likely, AGB
variables.
Solid line indicates the  mean  locus for K giants
\citep{frogel78}.
Dotted and dashed lines delimit the CMD region
occupied by field LPV and Carbon stars, 
solid line the mean locus of field K giants 
\citep[][ F95 and references therein]{bb88}.
\label{figure6}}
\end{figure}

Figure \ref{figure6} shows the 
cumulative $\rm (J-H)_0,(H-K)_0$ de-reddened 
color--color diagram for the six observed LMC clusters.
The majority of the brightest stars at $\rm K_0< 12.3$ 
have very red $\rm (J-H)_0$ and $\rm (H-K)_0$ colors, 
typical of long period variables and/or carbon stars 
which are the brightest extension of the AGB
evolutionary stage, usually significantly brighter
that the RGB-Tip.
This is a further confirmation that our RGB Tip assumption 
($\rm K_0\approx 12.3$) is correct.

\section{Star counts and integrated luminosities}
\label{counts}

In each cluster the observed stars have been classified as 
AGB, RGB and He-clump stars, 
accordingly to the {\it selection boxes}
defined in Section \ref{cmdsec} 
and sketched in Figure \ref{figure5}. 
 
In order to obtain reliable stellar counts and integrated luminosities 
in each evolutionary sequence, 
it is also necessary to 
estimate the degree of completeness of the sample and its 
possible contamination by the foreground/background stars and to 
correct for them.

\subsection{Completeness}
\label{compl}

Completeness has been estimated by using the well known
artificial star techniques. Here, we briefly recall the
main steps.  The RGB fiducial line for each cluster was
derived,   then a population of artificial stars, having
magnitudes, colors and luminosity function resembling those
of the observed RGB distribution was generated and added to
the original images. Only a sub-raster centered on the
cluster and included within a radius  of 1'.5$\times$1'.5
was used to estimate completeness. Outside this area
crowding effects are negligible and the completeness level
is $\approx 100\%$.

A total of 100,000 stars were simulated with a distribution
resembling the observed one.  The fraction of recovered objects
was estimated as \( \Lambda =N_{rec}/N_{sim} \). The observed
distribution is in principle distorted because of two phenomena: the loss of
faint stars due to incompleteness and an excess of bright stars
due to possible blending of two or more faint stars into a brighter one.
The first effect
is taken into account by the artificial star simulation. The second
is more complicated and it can be evaluated by analyzing the  
histogram of the magnitude difference between input and recovered
magnitudes. It turns to be negligible
\citep[see also discussion in][ and references therein]{testa99}. 

\begin{figure}
\figurenum{7}
\plotone{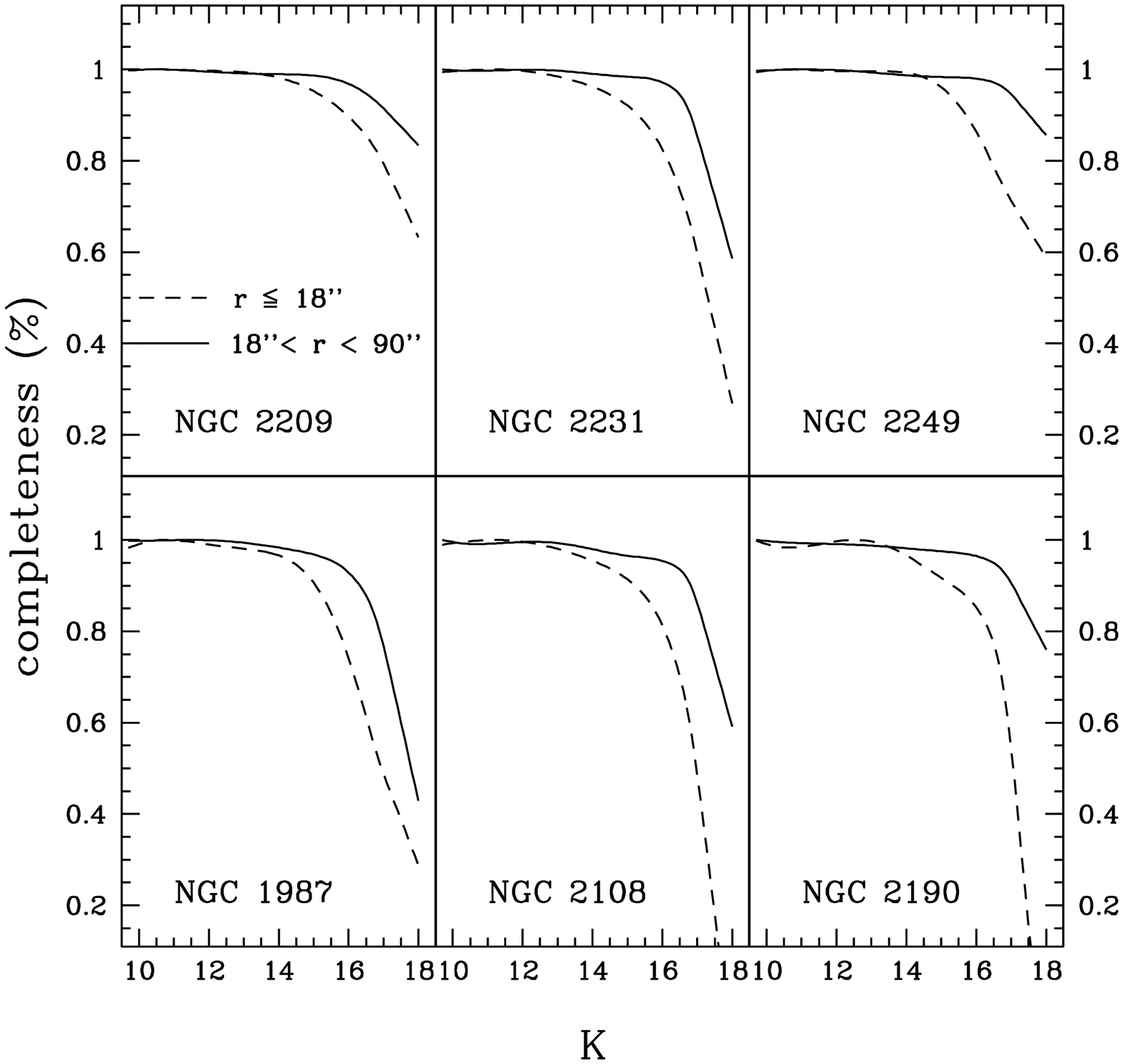}
\caption{Completeness curves. 
Dashed lines: $r\le18\arcsec$, solid lines: $18\arcsec
<r<90\arcsec$ 
(see Section \ref{compl}).
\label{figure7}}
\end{figure}

Figure \ref{figure7} shows  
the curves of completeness for the target clusters   
in two different annuli, namely r$\le$18"  and r$>$18''  
within the selected 1.5'$\times$1.5' area. 
At $\rm K<14$ completeness is always $>$95\%.
At $\rm K\approx 17$ completeness is $\ge $80\% in the outer annulus 
and $\ge $50\% in the inner one where crowding is more 
severe.
The cutoff radius which divides the two annuli
was set on the basis of the radial stellar distribution.  
However, its choice is not critical since 
reasonable small variations (within $\approx$20\%) 
from our best-fiducial value turns out to 
have negligible effects on the results.

\subsection{Field de-contamination}

The issue of de-contaminating the observed 
star distribution from the field contribution 
is rather tricky. 
The method usually employed is known as {\it zapping technique}: 
the CMDs of the {\it contaminated} cluster 
and the field  populations 
are divided in cells of fixed magnitude and color widths. 
After normalizing the areas, 
the {\it cleaned} CMD is obtained by randomly subtracting, cell by cell,
the number of stars found in the field CMD from the {\it contaminated} cluster
one. 
This technique has some drawbacks which are described in detail
in various papers 
\citep[see e.g. the discussion in][ and references therein]{testa99}.
In our case, the main limitation is the relatively low number 
statistics of some cells, which can generate a {\it cleaned} CMD with a   
patchy distribution and entire zones artificially de-populated. 

We thus preferred to directly de-contaminate the star counts.
The stars in the inner region (within a radius of 1.5' 
from the cluster center) 
are considered  as representative of the cluster 
{\it contaminated} population, while 
those in the outermost regions  (at a radius of 
$\gtrsim 2'$ from the cluster center)  
as representative of the field population. 

The total number of stars observed in each evolutionary
sequence  (AGB, RGB and He-clump) has been counted
accordingly to the {\it selection boxes} 
shown in Figure \ref{figure5} both in the 
{\it contaminated} cluster and field CMDs.
The star counts from the {\it contaminated} cluster population 
have been then corrected for
incompleteness adopting the curves shown in Figure \ref{figure7}.
The star counts in the field population have been
scaled to take into account the different surveyed area, 
and their contribution has been subtracted from the cluster 
{\it contaminated} population. At the end of this procedure  
{\it cleaned} LFs for each evolutionary sequence and in each cluster 
have been finally obtained.
  
Table \ref{tab3} list the final star counts in each sequence 
for the de-contaminated cluster and field populations.
As already mentioned, two clusters 
in our sample, namely NGC~1987 and NGC~2108, have 
a significant degree of field contamination,  
$\approx$46 and 41\% by number, respectively.
For the other clusters the degree of field contamination is 
much lower (between 10 and 24\%).

\section{Population ratios and evolutionary timescales}
\label{poprat}

When dealing with resolved stellar populations a number of interesting tests 
of theoretical models of stellar evolution can be performed. 
In this respect, population ratios (by numbers and/or by luminosity) are  
crucial tools to calibrate the relative lifetimes 
(and luminosity contribution) of each specific 
evolutionary sequence, to identify special 
events like the {\it phase transitions},  
and to check the overall reliability of the stellar clock. 

\subsection{Definition of the observables}    

As mentioned in Section 1, F95 presented preliminary observational
evidences that the RGB {\it Ph--T} does occur at $s=35-36$, however
due to the modest performances of the IR detector used at that time
the analysis was limited to the brightest 2-mag bin
of the RGB ($\rm 12.3<K_0<14.3$). 
In order to  account for the size
of the total cluster population,
both star counts and luminosity
of the upper RGB in F95 
were  normalized to the cluster integrated
luminosity \citep[in unit of $10^4 L_{\odot}$ taken from][]{perss83}. 
The adopted normalization was the
only possible one, due to the limited extension
in magnitude of the CMD. 
However, it should be noted that the integrated luminosity 
of intermediate-age clusters can be   
easily dominated by the few bright AGB stars, hence it can be   
significantly affected by stochastic effects due to the 
intrinsic low number statistics characterizing the AGB evolutionary phase 
\citep[][ see also Figure~11 in F95]{chiosi86,sf97}. 

An alternative {\it full empirical} 
normalization, less affected by stochastic effects, 
can be obtained  by using the He-clump population.
All the intermediate-age and old stars are indeed expected to experience  
a stable He core-burning phase \citep[see e.g.][ for a review]{osm96}, 
whose signature in the CMD is a red clump about 
$\rm \delta K\approx  5$~mag below the RGB Tip.  
The He-clump is clearly visible in all the CMDs 
shown in Figures \ref{figure2} and \ref{figure3}, hence 
two observables   
(which can be directly estimated from the CMD)
can be defined: the population ratio 
between the RGB and He-clump ($N_{RGB}/N_{He-C}$) number of stars and their 
corresponding luminosities ($L^{bol}_{RGB}/L^{bol}_{He-C}$). 

Global luminosities for each evolutionary stage
have been computed by summing the contribution of  
the single stars in the completeness-corrected and 
field de-contaminated LF obtained for each
{\it selection box} defined in Section \ref{cmdsec} and in Figure \ref{figure5}.
In particular, the de-reddened
K-band magnitudes have been scaled to absolute values    
by adopting a distance modulus $\rm (m-M)_0=18.5$ \citep{vdb98,nw00} and 
converted to bolometric magnitudes by 
applying suitable bolometric corrections  
(as derived by using 
the $\rm (J-K)_0$ color and the empirical calibrations by
\citet{monte98}.
The  derived  {\it integrated} bolometric luminosities  
(in unit of  $10^4 L_{\odot}$)
for each evolutionary stage (AGB, RGB,i He-clump) are
listed in Table 3.

\subsection{Theoretical models}
  
Theoretical predictions for the observables defined in the
previous section have been computed by using SSP models by \citet{maraston98} and 
\citet{maraston01}, for which the synthetic colors have been
calibrated on the observed integrated colors 
of MC clusters.

The adopted evolutionary code estimates 
the energetics of any post-MS phase
by using the so-called {\it Fuel Consumption Theorem}
\citep{rb86} and allows to model 
the two key AGB and RGB {\it phase transitions}. 
 
The main synthetic ingredients which mostly influence 
the theoretical predictions  are:
 
{\it (i)} the adopted  stellar evolutionary tracks. 
The stellar tracks used here are taken from \citet{cs97,bono97}. 
These are {\it canonical} tracks, without {\it overshooting}, in which
the most recent input physics (opacities, equation of state, etc.) are
adopted.  The mixing-length parameter has been calibrated on the Sun 
and scaled to other metallicities by using empirical relations \citep{sc96}.

{\it (ii)} the integration method.
The method adopted to determine the number of stars (and
luminosity) in any post-MS phase in the {\it Fuel Consumption} approach
is different with respect to that used in the isochrone 
technique, which is based on the mass dispersion along the post-MS
phases. Conversely here, the post-MS
stellar track of a mass equal to the Turnoff mass at a given stellar
population age is divided into a suitable 
number of sub-phases.  
Then, the  evolutionary timescale is combined with the Fuel Consumption in
order to evaluate  the number of stars and their luminosity in each sub-phase 
\citep[more details can be found in]{maraston98,maraston01}.

{\it (iii)} the temperature-color transformations.  
Transformations are taken from the
BaSel tables \citep{l97}, in which the
classical Kurucz library down to 3500~K is linked to models for cooler
temperatures \citep{bessel89}, and re-calibrated on
observed colors of individual stars. 

In order to make a preliminary check of the impact of the
different treatment of mixing on the observables described
in Section 5.1,  we have computed SSP models with the
procedure outlined in this section, but adopting the
stellar tracks with {\it overshooting} from Girardi et al.
(2000). We have chosen these tracks among the several
including a parameterization of overshooting, since in F95
we  used an earlier release of the Padova tracks. In a
future work we will explore other sets of stellar tracks
that do include an overshooting effect, but with different
efficiency (see e.g. \citet{yi01}).  Indeed, as quoted by
\citet{gallart03} the extension and efficiency of
overshooting is still not well established, and the
comparison with observations is crucial.

\subsection{Age calibration}                                         
\label{age}    

Since in the literature an homogeneous set of age
determinations based on the MS-TO is still lacking
for MC clusters, we have used the so-called {\it s}-parameter.
This parameter  was defined by \citet{ef85}        
as a curvilinear coordinate running along the mean locus
defined by MC clusters in the $(U-B,B-V)$ integrated
color-color diagram and it turns out to vary  
from 1 (very young clusters) to 51
(very old cluster). 

Of course, the absolute age for the programme clusters
obtained via the {\it s}-parameter
strictly depends on the adopted calibration,
which, in turn, 
depends on the  details of the theoretical models of stellar
evolution and, in particular, on the treatment of the convection. 
We adopted two different calibrations  available in
the literature:

\noindent
1) the  calibration 
obtained by Elson \& Fall (1988, hereafter EF88) 
based on {\it canonical} models:
$$log~t=6.05 + 0.079 \times s$$
2) the  {\it  overshooting} calibration
presented by \citet{girardi95} and based on 
the models with {\it overshooting} by \citet{b94}:
$$log~t=6.227 + 0.0733 \times s$$ 

Table~4 lists the ages of the six clusters in our sample as 
obtained from the {\it s}-parameter, using the two
above calibrations (see column 3 and 4).
Surprisingly, the two calibrations provide
very similar ages (within 10\%), hence
in the following we adopt  
the most recent one by Girardi et al. (1995), 
finding that the target clusters span an age range  
between $\approx$500 and 900~Myr.
Table~4 also lists the few age determinations
available in the literature from direct
measurements of the MS-TO. 
Ages determined through  the {\it s}-parameter calibrations 
agree with the TO ones within 20--30\%. 
Hence one can safely conclude that the {\it s}-parameter provides 
a reasonable estimate (within $\pm \simeq $20\%) of the cluster age.
According with this result, a conservative estimate of
the error ($\approx 20\%$) in the age determination has
been assigned to each cluster, this value has been
computed assuming an uncertainty of $\delta s=\pm1$, which
correspond to  $\delta(age)=\pm100$~Myr at $s=35$
($t\approx 620$~Myr).

\subsection{Comparison between observations and model predictions}
\label{compar}
 
\begin{figure}
\figurenum{8}
\plotone{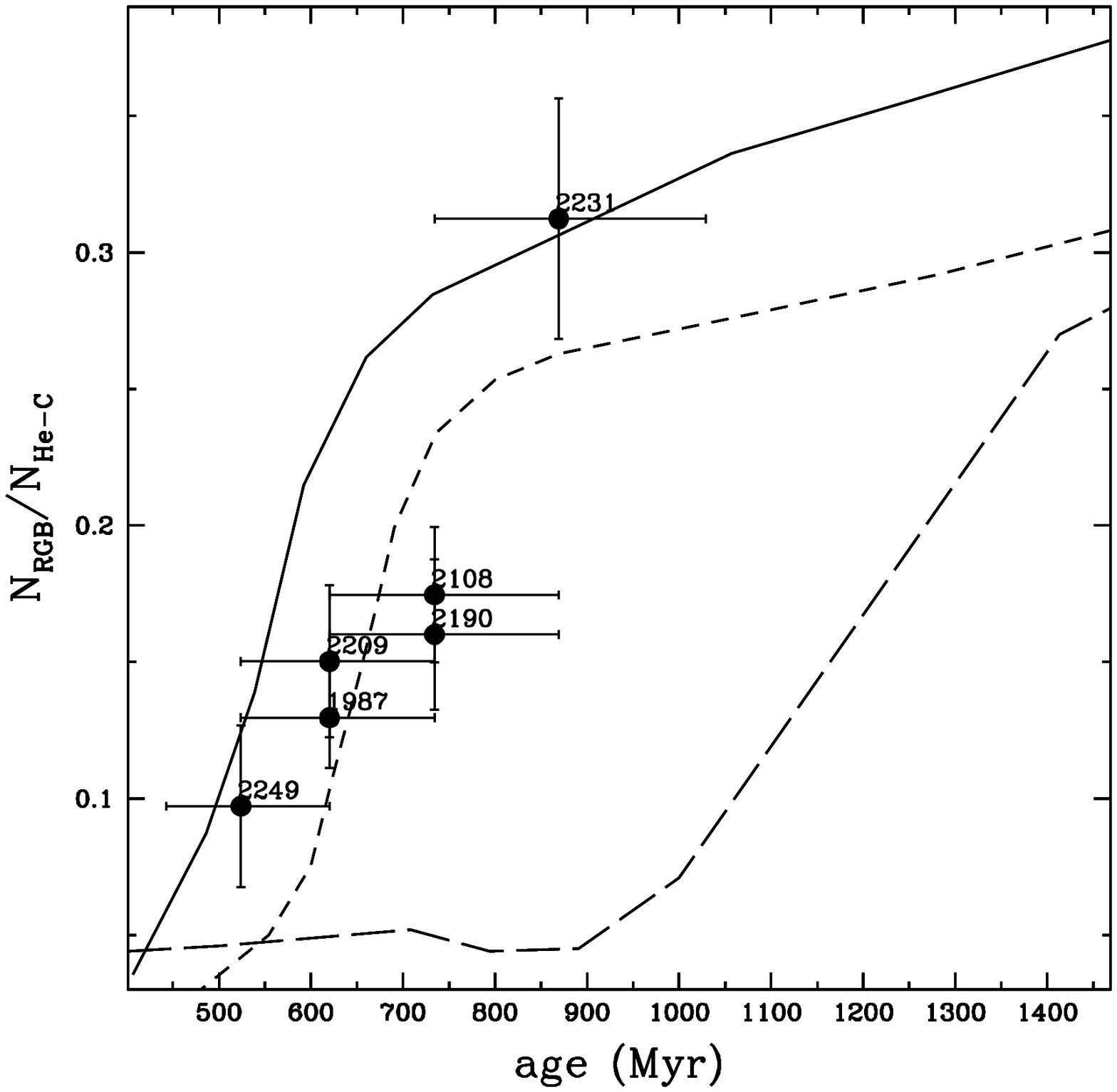}
\caption{
Ratio between the number of the bright RGB and He-clump 
stars as a function of age for the six observed clusters,
Stars  belonging to the   two populations are selected
accordingly  to the {\it selection boxes}     defined in
Section \ref{poprat} and shown  in Figure \ref{figure5}. 
Short-dashed and solid lines represent {\it canonical} models at
two different metallicities,  namely [Z/H]=--0.33 and
--1.35, respectively,   \citep[see][ and Section
\ref{counts}]{maraston98}, while long-dashed line is  the
result by using  {\it overshooting}  tracks at [Z/H]=--0.4 from
\citet{girardi00}.
\label{figure8}}
\end{figure}

\begin{figure}
\figurenum{9}
\plotone{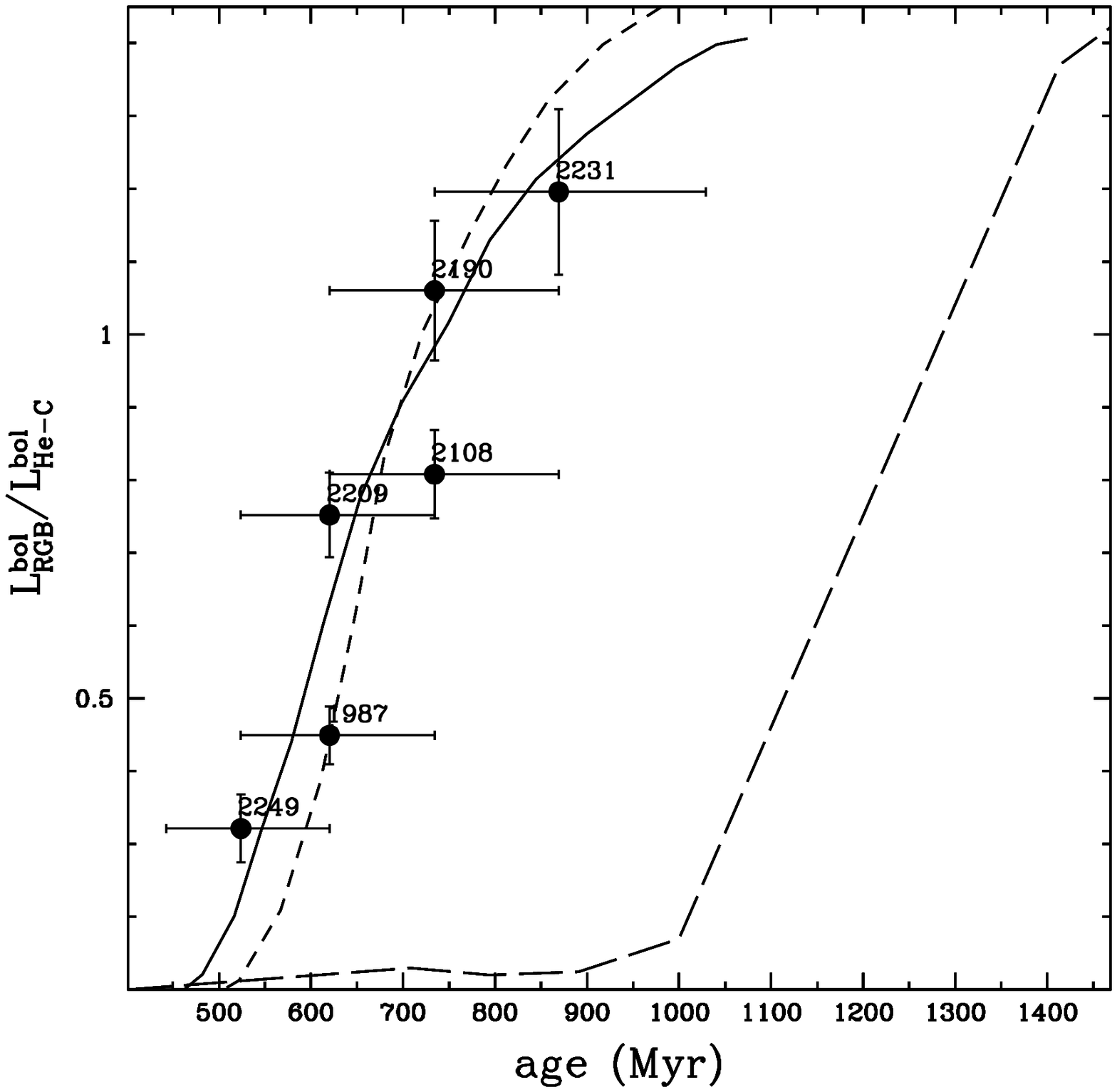}
\caption{
Ratio between the bolometric luminosity of the bright RGB
and He-clump  stars as a function of age for the six
observed clusters. Star belonging to the  two populations
are selected accordingly to the {\it selection boxes}
defined in Section \ref{poprat} and shown  in Figure
\ref{figure5}. Short-dashed and solid lines represent
{\it canonical} models at two different metallicities,  namely
[Z/H]=--0.33 and --1.35, respectively \citep[see ][ and
Section \ref{counts}]{maraston98},  while long-dashed line
is the result by using  {\it overshooting} tracks at [Z/H]=--0.4 from
\citet{girardi00}.
\label{figure9}}
\end{figure}

Figure \ref{figure8} shows the ratio between the 
number of RGB and the He-clump stars in each cluster, 
according to the selection criteria illustrated in Section \ref{cmdsec} 
and in Figure \ref{figure5},
as a function of the cluster age.
Figure \ref{figure9} shows the ratio between  
the corresponding  bolometric luminosities.
By inspection of Figure \ref{figure8} and \ref{figure9}
one can see that the contribution
of the RGB phase increases by a factor of 
$\approx$3 by number (cf. Figure \ref{figure8} and Table \ref{tab3}) and  
$\approx$4 by luminosity (cf. Figure \ref{figure9} and Table \ref{tab3})
in less than $\approx 400$~Myr.

Both Figures 8 and 9 support the hypothesis that the
cluster set presented here properly sample
the epoch of the full development 
(both in luminosity and in star number)
of the RGB.
This result fully confirms the finding by F95
who identified NGC1987 and NGC2108 as the two intermediate-age 
clusters on the verge of the RGB {\it Ph--T}.
 
The lines in Figure \ref{figure8} and \ref{figure9}
represent model predictions: short-dashed and solid lines, 
{\it canonical} models 
\citep[][ see also Section 5.2]{maraston98}, 
at two different metallicities, 
namely [Z/H]=--0.33 and --1.35, respectively,
long-dashed line model with {\it overshooting} \citep{girardi00}
at [Z/H]=--0.4 and Y=0.25, the most representative for the LMC
\citep[see e.g.][]{gb98}. 

 The time-scale yielded by the two calibrations of the 
{\it s}-parameter discussed in Section 5.1 shows a nice agreement
with that defined by the  {\it canonical} models, suggesting that
the RGB {\it Ph--T} occurs at $t\approx 700$~Myr and it is a 
quite rapid event since the main increase of the RGB luminosity 
occurs in less than $\delta t \approx 300$~Myr.
 
Both {\it overshooting} and {\it canonical} models agree
with  observations in showing that the RGB {\it Ph--T}  is
a rapid event ($\delta t \approx 300$~Myr).  However,
evolutionary tracks with {\it overshooting}  predict the
RGB {\it Ph--T} occurrence at significantly later  ($\Delta
t\simeq 500$~Myr) epochs than the one  indicated by {\it
canonical} models. Indeed, by carefully inspecting  the
Girardi et al. (2000) isochrones, one finds  that at 
t=1~Gyr  the RGB-tip luminosity is still steeply
increasing:   at $[M/H]=-0.4$ the absolute magnitude of the
RGB-Tip ($M_V(RGB-tip)$) changes from $M_V(RGB-tip)=-0.71$
at t=1 Gyr to $M_V(RGB-tip)=-2$ at t=1.4 Gyr.  This
confirms that at the epoch ($t\approx 1$ Gyr) at which the
RGB {\it Ph--T} is almost completed in the {\it canonical}
models (see Figure 8 and 9), it is still  under development
when using the {\it overshooting} evolutionary tracks.

The mismatch shown in Figures 8 and 9 suggest some problems
with the evolutionary timescales  of the {\it overshooting}
models by \citet{girardi00} and/or  with the
\citet{girardi95} calibration of the {\it s}-parameter. 
However, a similar discrepancy was already noted by F95 in
the previous generation of {\it overshooting} models. \\ A
new calibration of the {\it s}-parameter in terms of age by
using high quality CMDs and updated models is urgently 
needed to clarify this issue.

\section{Summary and future work}                  

In  this paper we  present the photometric 
analysis based on near IR CMDs down to $\rm K\approx 18.5$ of the 
six intermediate-age
LMC clusters, namely NGC~1987, NGC~2108,
NGC~2190, NGC~2209, NGC~2231 and NGC~2249, in our sample.   
More than 4,000 giant stars in the cluster and surrounding field populations 
have been studied.
 
As a suitable diagnostics to probe the existence of the RGB {\it Ph--T} 
we use the ratios (by number and luminosity)
between the RGB and He-clump populations.
Empirical estimates have been compared with model predictions.
The observed RGB population shows a sharp enhancement 
(both in number and luminosity) at $s=36$.
 This corresponds to $t\approx 700$~Myr, 
accordingly with the current available calibrations
of the {\it s}-parameter (both using {\it canonical} and 
{\it overshooting} models). 
However, in the case of models with overshooting, 
a significant discrepancy between the timescale provided 
by the {\it s}-parameter and the one set by the evolutionary tracks 
has been found.
The origin of such a discrepancy needs to be further investigated.
 
In forthcoming papers we will present the near IR CMDs for
the other MC clusters in our sample, spanning the full 
age range between $10^8$ and a few $10^9$ yrs. 
 The analysis of the entire sample
will allow us to fully  characterize the photometric
properties of the AGB, RGB
and He-clump sequences with varying the age of the stellar
population.

\acknowledgments
The financial support by the Agenzia Spa\-zia\-le Ita\-lia\-na (ASI)
and the Ministero dell'Istru\-zio\-ne, Universit\`a e Ricerca (MIUR)
is kindly acknowledged.

\clearpage

\begin{deluxetable}{llllll}
\tablecolumns{6} 
\tablewidth{0pc} 
\tablecaption{Main parameters of the target clusters.\label {tab1}} 
\tablehead{ 
\colhead{cluster}   &  
\colhead{$\alpha $} &   
\colhead{$\delta $} & 
\colhead{\emph{s}} &
\colhead{[Fe/H]}   & 
\colhead{E(B-V)}} 
\startdata
NGC~1987 &05:27:17 &-70:44.1 &35 &--1.00 &0.12\\
NGC~2108 &05:43:56 &-69:10.8 &36 &--1.20 &0.18\\
NGC~2190 &06:01:02 &-74:43.5 &36 &--0.12 &0.10\\
NGC~2209 &06:08:34 &-73:50.5 &35 &--1.20 &0.07\\
NGC~2231 &06:20:44 &-67:31.1 &37 &--0.67 &0.08\\
NGC~2249 &06:25:49 &-68:55.2 &34 &--0.12 &0.10\\
\enddata 

\tablecomments{$~~~$\\
{\it s}-parameter from \citet{ef85} and \citet{girardi95}.
Metallicity and reddening from \citet{corsi94} and F95 
compilations. 
For NGC~2231 not included in \citet{corsi94} and F95 sample, metallicity  
is from \citet{olsze91} and 
reddening from \citet{perss83}.} 
\end{deluxetable} 


\begin{deluxetable}{rrrrrrrrr} 
\tablecolumns{9} 
\tablewidth{0pc} 
\tablecaption{Photometry of the target clusters \label {tab2}} 
\tablehead{ 
\colhead{Id} & \colhead{X}   & \colhead{Y}    & \colhead{J} & 
\colhead{$\sigma$(J)}    & \colhead{H}   & \colhead{$\sigma$(H)}    & 
\colhead{K}   & \colhead{$\sigma$(K)}}
\startdata 
\emph{NGC~1987} & & & & & & & & \\
    1 &  740.819 & 851.305&  9.819&  0.011&  9.558&  0.007&  9.519&  0.007 \\
    2 &  501.969 & 368.297& 12.102&  0.007& 10.933&  0.014& 10.222&  0.014 \\
    3 &  363.867 & 652.862& 12.146&  0.002& 11.001&  0.012& 10.341&  0.012 \\
    4 &  491.525 & 460.867& 12.008&  0.005& 11.088&  0.019& 10.768&  0.019 \\
    5 &  321.522 & 563.854& 12.031&  0.007& 11.098&  0.011& 10.788&  0.011 \\
    6 &  746.735 & 480.059& 12.540&  0.006& 11.552&  0.020& 11.030&  0.020 \\
    7 &  483.386 & 461.947& 12.497&  0.009& 11.611&  0.022& 11.275&  0.022 \\
    8 &  909.377 & 911.159& 12.588&  0.003& 11.641&  0.015& 11.365&  0.015 \\
    9 &  597.747 & 432.049& 12.659&  0.006& 11.741&  0.017& 11.426&  0.017 \\
   10 &  390.194 & 379.146& 13.137&  0.007& 12.202&  0.017& 11.905&  0.017 \\
\enddata 
\end{deluxetable} 


\begin{deluxetable}{lrrrrrrr}   
\tablecolumns{9} 
\tablewidth{0pc} 
\tablecaption{Star counts and bolometric luminosities$^a$. \label {tab3}} 
\tablehead{ 
\colhead{cluster}             & 
\colhead{$\rm N_{AGB}$}       & 
\colhead{$\rm N_{RGB}$}       & 
\colhead{$\rm N_{He-C}$}  & 
\colhead{~~~}                 & 
\colhead{$\rm L_{AGB}^{bol}$} & 
\colhead{$\rm L_{RGB}^{bol}$} & 
\colhead{$\rm L_{He-C}^{bol}$} }  
\startdata 
\underline{\emph{cluster population$^b$}}  & & & & & & & \\
    NGC~1987   &   6   &   42  &   322   &&   3.65   &  0.92   &  2.05 \\
    NGC~2108   &   3   &   40  &   231   &&   2.72   &  1.11   &  1.38 \\
    NGC~2190   &   2   &   28  &   174   &&   0.95   &  0.94   &  0.89 \\
    NGC~2209   &   3   &   24  &   160   &&   1.53   &  0.61   &  0.81 \\
    NGC~2231   &   1   &   36  &   114   &&   0.46   &  0.71   &  0.59 \\
    NGC~2249   &   1   &    9  &    98   &&   0.32   &  0.16   &  0.51 \\       
\underline{\emph{field population$^c$}} & & & & & & & \\
    NGC~1987   &   3   &   90  &   222   &&   0.68   &  2.83   &  1.26 \\
    NGC~2108   &   2   &   63  &   124   &&   0.82   &  1.94   &  0.68 \\
    NGC~2190   &   0   &    7  &    18   &&   0.00   &  0.17   &  0.11 \\
    NGC~2209   &   0   &   12  &    34   &&   0.00   &  0.28   &  0.20 \\
    NGC~2231   &   0   &   13  &    21   &&   0.00   &  0.33   &  0.14 \\
    NGC~2249   &   0   &   11  &    29   &&   0.00   &  0.45   &  0.17 \\      
\enddata 
\tablecomments{$~~~~$\\
$^{\bf a}$~Star counts  
  are corrected for incompleteness
(see Section \ref{compl}).
Bolometric luminosities are in units of $\rm 10^4~L_{\odot}$.\\
$^{\bf b}$~The cluster population is de-contaminated by the
field one.\\
$^{\bf c}$~The field population 
has been normalized to the $\rm \approx7~arcmin^2$ area,  
as sampled by the cluster population (see Section \ref{cmdsec} for details).}
\end{deluxetable} 

\begin{deluxetable}{llllll}
\tablecolumns{6} 
\tablewidth{0pc} 
\tablecaption{Age determinations for the target clusters.\label {tab4}} 
\tablehead{ 
\colhead{cluster}   &  
\colhead{\emph{s}$^a$} &
\colhead{\emph{s}--age$^b$} &
\colhead{\emph{s}--age$^c$} &
\colhead{TO--age}$^d$ &
\colhead{Reference$^e$} }
\startdata
NGC~1987 &35 &653 &620&479    &\citet{girardi95}\\
NGC~2108 &36 &783 &734&525    &\citet{girardi95}\\
NGC~2190 &36 &783 &734&813    &\citet{girardi95}\\
NGC~2209 &35 &653 &620&800    &\citet{g76}\\
         &   &    &&700    &\citet{hodge83}\\      
         &   &    &&910    &\citet{girardi95}\\
NGC~2231 &37 &940 &869&1200   &\citet{hodge83}\\   
NGC~2249 &34 &544 &524&550$^f$    &\citet{v94}\\
         &   &    &&300$^g$    &\citet{v94}\\
\enddata 
\tablecomments{$~~~$\\
($^a$)~{\it s}-parameter from \citet{ef85} and \citet{girardi95}.\\
($^b$)~Age in Myr, according to the calibration by EF88: 
$log~t=6.05 + 0.079 \times s$.\\
($^c$)~Age in Myr, according to the calibration by \citet{girardi95}: 
$log~t=6.227 + 0.0733 \times s$.\\
($^d$)~TO age in Myr.\\ 
($^e$)~Reference for the TO-age determination.\\
($^f$)~TO-age from models with {\it overshooting} \citep{v94}.\\
($^g$)~TO-age from models with {\it no-overshooting} \citep{v94}.} 
\end{deluxetable} 

\clearpage
 
\end{document}